%% file: etapipi.tex
\documentclass[aps,prd,reprint,amsmath,amssymb,showpacs,superscriptaddress,nofootinbib,twocolumn]{revtex4-1}
\usepackage{multirow}
\usepackage{amssymb}
\usepackage{graphicx}
\usepackage{graphics}
\usepackage{epsfig}
\usepackage{bigstrut}
\usepackage{bm}
\usepackage{xspace}
\usepackage{amsmath}
\usepackage[mathlines]{lineno}
\usepackage{color}
\usepackage[bookmarksnumbered, pdfstartview=FitH,colorlinks,citecolor=blue,linkcolor=blue,]{hyperref}
\usepackage[perpage,symbol]{footmisc}
\usepackage{subfigure}
\usepackage{indentfirst}
\usepackage{fancyvrb}
\usepackage{float}
\usepackage{textpos}
\usepackage[english]{babel}

\newcommand{\ra}{\rightarrow}

\newcommand{\jpsi}{J/\psi}
\newcommand{\pio}{\pi^{0}}
\newcommand{\pip}{\pi^{+}}
\newcommand{\pim}{\pi^{-}}
\newcommand{\etap}{\eta^{\prime}}
\newcommand{\chisq}{\chi^{2}}

\renewcommand{\thefootnote}{\fnsymbol{footnote}}
\begin{document}

\title{Measurement of the matrix elements for the decays $\etap\ra\eta\pip\pim$ and $\etap\ra\eta\pio\pio$ }

\input{authors}

\begin{abstract}
Based on a sample of $1.31\times10^9$ $\jpsi$ events collected with the BESIII detector,
the matrix elements for the decays $\etap \to \eta \pip\pim$ and
$\etap \to \eta\pio \pio$ are determined using 351,016 $\etap\ra(\eta\to\gamma\gamma)\pip\pim$ and 56,249
$\etap\ra(\eta\to\gamma\gamma)\pio\pio$ events with background levels less than 1\%.
Two commonly used representations are used to describe the Dalitz plot density.
We find that an assumption of a linear amplitude does not describe the data well. 
A small deviation of the obtained matrix elements 
between $\etap\ra\eta\pip\pim$ and $\etap\ra\eta\pio\pio$ 
is probably caused by the mass difference between charged and neutral pions
or radiative corrections.  
No cusp structure in $\etap\ra\eta\pio\pio$ is observed.
\end{abstract}


\maketitle
\input{main}

\begin{acknowledgments}
The BESIII collaboration thanks the staff of BEPCII and the IHEP computing center for
their strong support.
This work is supported in part by National Key Basic Research
Program of China under Contract No. 2015CB856700; National Natural Science Foundation
of China (NSFC) under Contracts No. 11235011, No. 11335008, No. 11425524,
No. 11625523, No. 11635010, No. 11675184, No. 11735014;
the Chinese Academy of Sciences (CAS) Large-Scale Scientific Facility Program;
the CAS Center for Excellence in Particle Physics (CCEPP);
Joint Large-Scale Scientific Facility Funds of the NSFC and CAS under
Contracts No. U1332201, No. U1532257, No. U1532258; CAS under Contracts
No. KJCX2-YW-N29, No. KJCX2-YW-N45, No. QYZDJ-SSW-SLH003; 100 Talents Program of CAS;
National 1000 Talents Program of China; INPAC and Shanghai Key Laboratory
for Particle Physics and Cosmology; German Research Foundation DFG under
Contracts No. Collaborative Research Center CRC 1044, No. FOR 2359;
Istituto Nazionale di Fisica Nucleare, Italy; Joint Large-Scale Scientific
Facility Funds of the NSFC and CAS; Koninklijke Nederlandse Akademie van
Wetenschappen (KNAW) under Contract No. 530-4CDP03; Ministry of Development
of Turkey under Contract No. DPT2006K-120470; National Natural Science
Foundation of China (NSFC); National Science and Technology fund;
The Swedish Research Council; U. S. Department of Energy under Contracts
No. DE-FG02-05ER41374, No. DE-SC-0010118, No. DE-SC-0010504, No. DE-SC-0012069;
University of Groningen (RuG) and the Helmholtzzentrum fuer
Schwerionenforschung GmbH (GSI), Darmstadt; WCU Program of
National Research Foundation of Korea under Contract No. R32-2008-000-10155-0.
We would like to thank Bastian Kubis for providing the
predicted distribution in dispersive analysis.
\end{acknowledgments}

\footnotemark[0]{Node added -- Recently, we notice that the A2 Collaboration also
presented a result on $\etap\ra\eta\pio\pio$~\cite{MAMI}.}

\bibliographystyle{unsrt}
\bibliography{etapipi}

\end{document}

%% file: authors.tex
\author{
 \begin{small}
 \begin{center}
M.~Ablikim$^{1}$, M.~N.~Achasov$^{9,d}$, S. ~Ahmed$^{14}$, M.~Albrecht$^{4}$, A.~Amoroso$^{53A,53C}$, F.~F.~An$^{1}$, Q.~An$^{50,40}$, J.~Z.~Bai$^{1}$, Y.~Bai$^{39}$, O.~Bakina$^{24}$
% ...
\\
\vspace{0.2cm}
(BESIII Collaboration)\\
\vspace{0.2cm} {\it
$^{1}$ Institute of High Energy Physics, Beijing 100049, People's Republic of China\\
% ...
$^{4}$ Bochum Ruhr-University, D-44780 Bochum, Germany\\
% ...
$^{9}$ G.I. Budker Institute of Nuclear Physics SB RAS (BINP), Novosibirsk 630090, Russia\\
% ...
$^{14}$ Helmholtz Institute Mainz, Johann-Joachim-Becher-Weg 45, D-55099 Mainz, Germany\\
% ...
$^{24}$ Joint Institute for Nuclear Research, 141980 Dubna, Moscow region, Russia\\
% ...
$^{39}$ Southeast University, Nanjing 211100, People's Republic of China\\
$^{40}$ State Key Laboratory of Particle Detection and Electronics, Beijing 100049, Hefei 230026, People's Republic of China\\
% ...
$^{50}$ University of Science and Technology of China, Hefei 230026, People's Republic of China\\
% ...
$^{53}$ (A)University of Turin, I-10125, Turin, Italy; (B)University of Eastern Piedmont, I-15121, Alessandria, Italy; (C)INFN, I-10125, Turin, Italy\\
% ...
$^{d}$ Also at the Novosibirsk State University, Novosibirsk, 630090, Russia\\
% ...
}
\end{center}
\vspace{0.4cm}
\end{small}
}

%% file: main.tex
\section{Introduction}\label{sec:introduction}

The $\eta^\prime$ meson is well
established and its main decay modes are fairly well known~\cite{PDG2014090001}.
However, $\eta^\prime$ decay dynamics remains a subject of extensive
theoretical studies aiming at extensions of the chiral perturbation theory (ChPT).
The two dominant hadronic decays, $\eta^\prime\rightarrow\eta\pi^+\pi^-$ and
$\eta^\prime\rightarrow\eta\pi^0\pi^0$ (called charged decay mode and
neutral decay mode throughout the text, respectively), are believed to be an ideal
place to study $\pi\pi$ and $\eta\pi$ scattering~\cite{PRD60034002,PhysRevD.90.033009}, which may lead to
a variation in the density of the Dalitz plot.
Several extensions of the ChPT framework~\cite{Beisert2003186,Borasoy2005383,Kubis:2009sb,Escribano2011094} and
dispersive analysis based on the fundamental principles of analyticity and unitarity~\cite{Isken2017} have been
applied to investigate the matrix element of
$\eta^\prime\rightarrow\eta\pi\pi$.

In experimental analyses, the Dalitz plot for the charged decay mode
is usually described by the following two variables:
\begin{equation}
	X=\frac{\sqrt{3}(T_{\pip}-T_{\pim})}{Q},\ \
	Y=\frac{m_{\eta}+2m_{\pi}}{m_{\pi}}\frac{T_{\eta}}{Q}-1.
\end{equation}
For the neutral decay mode, the Dalitz plot has twofold
symmetry due to the two $\pio$s in the final state. Hence, the variable
$X$ is replaced by
\begin{equation}
	X=\frac{\sqrt{3}|T_{\pi^{0}_{1}}-T_{\pi^{0}_{2}}|}{Q}.
\end{equation}
Here, $T_{\pi}$ and $T_{\eta}$ denote the kinetic energies of a pion and $\eta$ in
the $\etap$ rest frame, $Q=m_{\etap}-m_{\eta}-2m_{\pi}$, and $m_\pi$, $m_\eta$, and $m_\etap$ are
the masses of the pion, $\eta$, and $\etap$, respectively.
Generally, the decay amplitude squared is parametrized as
\begin{equation}\label{eq:ampxy}
\begin{gathered}
	|M(X,Y)|^{2}=N(1+aY+bY^2+cX+dX^2+\ldots),
\end{gathered}
\end{equation}
which is the so-called general representation. Here $a$, $b$, $c$ and $d$ are free
parameters and $N$ is a normalization factor.
The terms with odd powers in $X$ are forbidden due to the charge
conjugation symmetry in $\etap\ra\eta\pip\pim$ and the wave function symmetry in
$\etap\ra\eta\pio\pio$.
By considering the isospin symmetry,
the Dalitz plot parameters for the charged and neutral decay modes
should be the same. However, a small discrepancy, observed in previous
measurements~\cite{PRL8426, Dorofeev200722, PRD83012003, Blik2009231, Alde1986115}, is expected due to the mass
difference between the charged and neutral pion,
or due to radiative
corrections for the $\etap\ra\eta\pip\pim$ mode~\cite{Kubis:2009sb}.

A second parametrization for the decay amplitude squared used by previous experiments
assumes a linear amplitude in $Y$ and keeps
the polynomial expansion in $X$,
\begin{equation}\label{eq:ampxy_linear}
	|M(X,Y)|^2=N(|1+\alpha Y|^2 + cX+dX^2+\ldots),
\end{equation}
the so-called linear representation,
where $\alpha$ is a complex number. The real
part of $\alpha$ gives the linear term in $Y$ for the
Dalitz plot density, $a=2\Re(\alpha)$, and
the quadratic term is $b=\Re(\alpha)^2+\Im(\alpha)^2$,
where $\Re(\alpha)$ and $\Im(\alpha)$ are the real
and imaginary parts of $\alpha$, respectively.
The two representations are equivalent if $b>a^2/4$,
$i.e.$, $b$ should be at least larger than zero. Therefore,
a negative value for $b$ demonstrates that the ansatz of Eq.~(\ref{eq:ampxy_linear})
does not describe the data.

Experimentally, the decays of the $\etap\ra\eta\pip\pim$
and $\etap\ra\eta\pio\pio$ have only been explored with limited statistics so far.
The matrix elements for $\etap\ra\eta\pip\pim$ have been
studied by the CLEO (using only Eq.~(\ref{eq:ampxy_linear}))~\cite{PRL8426}, VES~\cite{Dorofeev200722} and
BESIII~\cite{PRD83012003} Collaborations. 
The most recent measurement
of $\etap\ra\eta\pio\pio$ is from the GAMS-4$\pi$
experiment~\cite{Blik2009231}, complementing older results reported by
the GAMS-2000 Collaboration~\cite{Alde1986115}.
Discrepancies in the Dalitz plot parameters both for the
charged and neutral decay channels are obvious from
those experiments.

In addition, the Dalitz plot for $\etap\ra\eta\pio\pio$ is expected to be affected by a cusp
due to the $\pip\pim$ mass threshold.
The size of this effect is predicted to be
about 6\%~\cite{Kubis:2009sbcorr} (8\% in original work~\cite{Kubis:2009sb}) within the framework of
nonrelativistic effective field theory (NREFT), which is confirmed in dispersive analysis~\cite{Isken2017}.
An analogous cusp has been observed in the
$K^+\ra\pip\pio\pio$~\cite{Batley2006173} decay and allows us to
determine the $\pi\pi$ $S$-wave scattering lengths
~\cite{Gamiz2007,PhysRevLett.93.121801,1126-6708-2005-03-021,Colangelo2006187}.

The dynamics of the decays
$\etap\ra\eta\pip\pim$ and $\etap\ra\eta\pio\pio$ are studied in this work using
$\etap$ mesons produced in the $\jpsi\ra\gamma\etap$ decay.
The present data sample of $1.31\times10^9$ $\jpsi$ events
accumulated with the BESIII detector is about 5 times of
that used in the previous BESIII analysis~\cite{PRD83012003}.

\section{Detector and Monte Carlo simulation}\label{sec:detector}

The BES\uppercase\expandafter{\romannumeral3} detector is a general-purpose
magnetic spectrometer with a geometrical acceptance of 93\% of 4$\pi$,
and is described in detail in Ref.~\cite{Ablikim2010345}.
The detector is composed of a helium-based drift chamber (MDC),
a plastic-scintillator time-of-flight system (TOF), and
a CsI(Tl) electromagnetic calorimeter (EMC), all enclosed in a
superconducting solenoidal magnet providing a 1.0~T magnetic field (0.9~T in 2012).
The solenoid is supported by an octagonal flux-return yoke with resistive-plate
counters interleaved with steel for muon identification (MUC).

The {\sc GEANT}4~\cite{Agostinelli2003250} based Monte Carlo (MC)
simulation software package {\sc BOOST}~\cite{dengzyboost} describes
the geometry and material of the BESIII detector, as well as
the detector response. It is used to optimize the event selection criteria,
estimate backgrounds and determine the detection efficiencies.
The production of the $\jpsi$ resonance is simulated with
{\sc KKMC}~\cite{PhysRevD63113009, Jadach2000260}, while the decays
are generated with {\sc EVTGEN}~\cite{Lange2001152,ChinPhysC.32.2008} for established modes
using world-average branching fractions~\cite{PDG2014090001},
and by {\sc LUNDCHARM}~\cite{PhysRevD62034003} for the remaining
decays.
An inclusive MC sample of $1.2\times10^9$ $J/\psi$ events is used
to study the potential background contributions.
The analysis is performed in
the framework of the BESIII off-line software system ({\sc BOSS})~\cite{wdliboss}.

\section{Measurement of the matrix element for $\etap\ra\eta\pip\pim$}\label{sec:charge}

For the reconstruction of $\jpsi\ra\gamma\etap$ with $\etap\ra\eta\pip\pim$
and $\eta\ra\gamma\gamma$, candidate events must contain two tracks
with an opposite charge and at least three photons. Each charged track
reconstructed from the MDC hits is required to have a polar angle in the range
$|\cos\theta|<0.93$ and to pass the interaction point within $\pm10$~cm
along the beam direction and within $\pm1$~cm in the plane perpendicular
to the beam. Photon candidates are reconstructed using isolated clusters
of energy deposited in the EMC and required to have a deposited energy
larger than 25~MeV in the barrel region ($|\cos\theta|<0.80$) or 50~MeV
in the end cap region ($0.86<|\cos\theta|<0.92$). The energy deposited
in nearby TOF counters is included to improve the
reconstruction efficiency and energy resolution. To eliminate
clusters associated with charged tracks, the angle between the photon
candidate and the extrapolation of any charged track to the EMC must be
larger than 10$^\circ$. A requirement on the EMC cluster timing with respect
to the event start time ($0 \le T \le 700$ ns) is used
to suppress electronic noise and energy
deposits unrelated to the event.

Since the radiative photon from the $\jpsi$ decay is always more energetic than
those from the $\eta$ decay, the photon candidate with the maximum energy
in an event is assumed to be the radiative one. For each
$\pip\pim\gamma\gamma\gamma$ combination, a six-constraint (6C) kinematic
fit is applied, and the $\chisq_{6C}$ is required to be less than 100.
The fit enforces energy-momentum conservation and constrains the invariant masses
of $\gamma\gamma$ and $\eta\pip\pim$ to the nominal $\eta$ and $\etap$
masses, respectively. 
If there are more than three selected photons in an event, the
combination with the smallest $\chi^{2}_{6C}$ is retained.

\begin{figure}
    \centering
    \includegraphics[height=6.5cm]{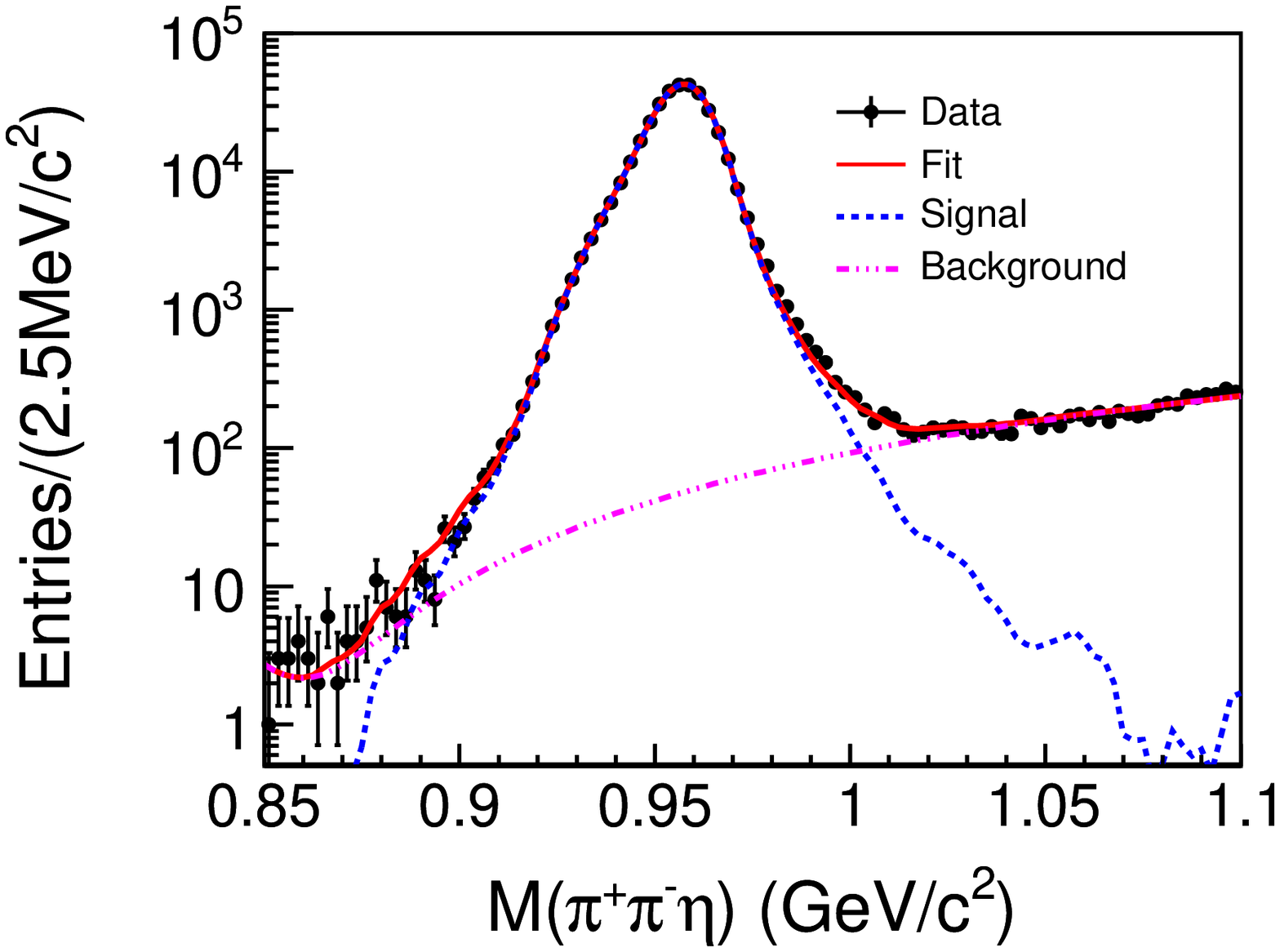}
    \caption{\label{fig:evtcha} Invariant mass spectrum of $\pip\pim\eta$ candidates
		without $\eta$ and $\etap$ mass constraints applied in the kinematic fit,
		and requiring the $\gamma\gamma$ invariant mass within the $\eta$ signal region.
		}
\end{figure}

To estimate the background contribution, an alternative data sample is selected
without applying the $\eta$ and $\etap$ mass constraints in the kinematic fit.
The $\pip\pim\gamma\gamma$ invariant mass spectrum
is shown in Fig.~\ref{fig:evtcha} after requiring the $\gamma\gamma$ invariant mass
within the $\eta$ signal region, (0.518, 0.578)~GeV/$c^2$.
A clear $\etap$ signal is observed with a low background level.
In addition, a sample of $1.2\times10^9$ inclusive MC $J/\psi$ decays
is used to investigate potential backgrounds.
Using the same selection
criteria for the MC sample, no peaking background remains around the $\etap$ signal region.
From this MC sample, the background contamination is estimated to be about 0.3\%.
This is consistent with an estimation obtained from an 
unbinned maximum likelihood fit to the $M(\eta\pip\pim)$ distribution,
where the signal is described by the MC simulated shape convoluted with a Gaussian
function representing the resolution difference between the data and MC simulation,
and the background contribution is described by a third-order polynomial function.
We therefore neglect the background
contribution in the determination of the Dalitz plot parameters.

\begin{figure}
\centering
		\includegraphics[height=6.5cm]{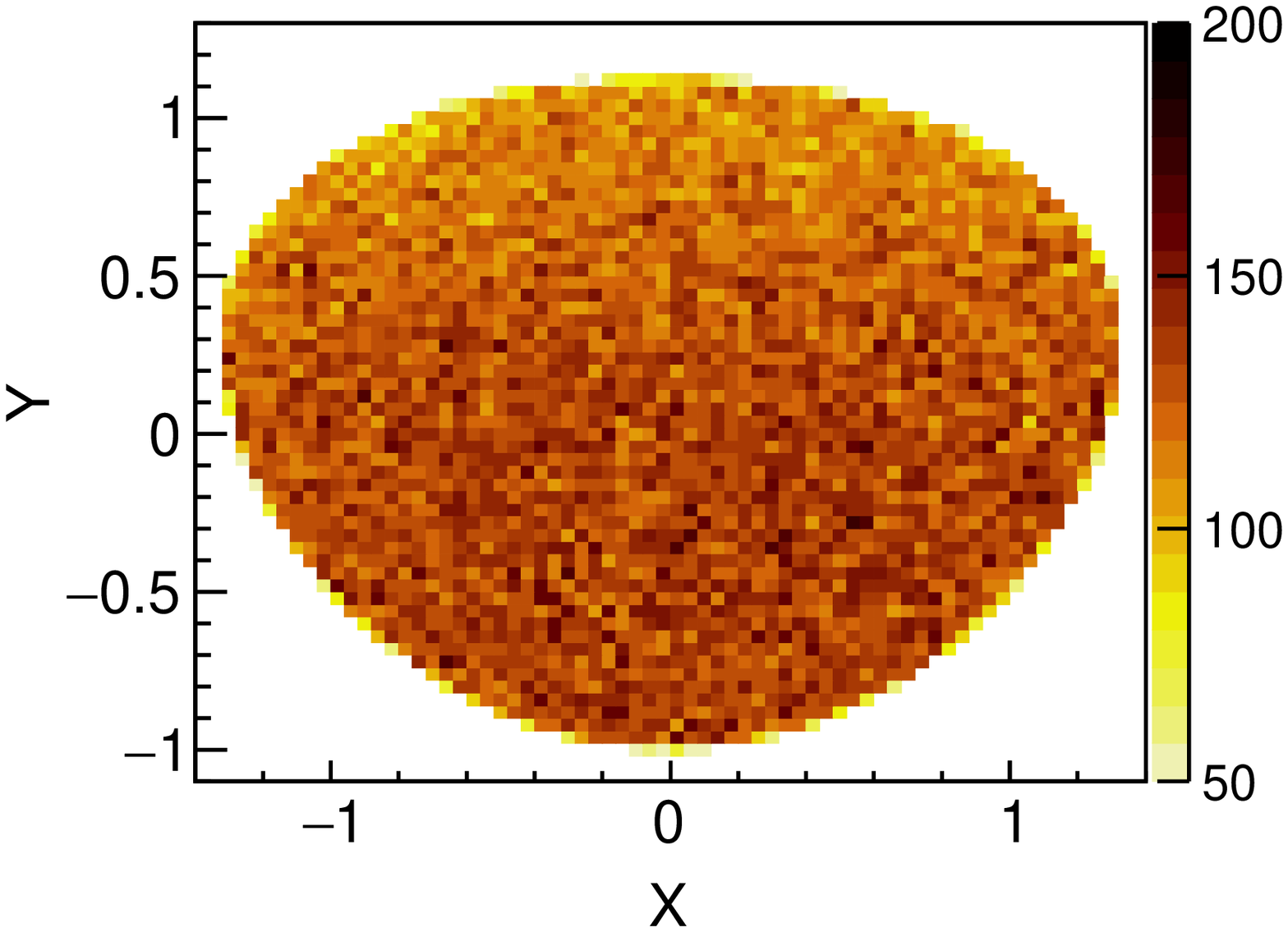}
    \caption{\label{Fig:ChaDalitz} Dalitz plot for $\etap\ra\eta\pip\pim$
		from data.}
\end{figure}

\begin{figure*}
\centering
		\includegraphics[height=8cm]{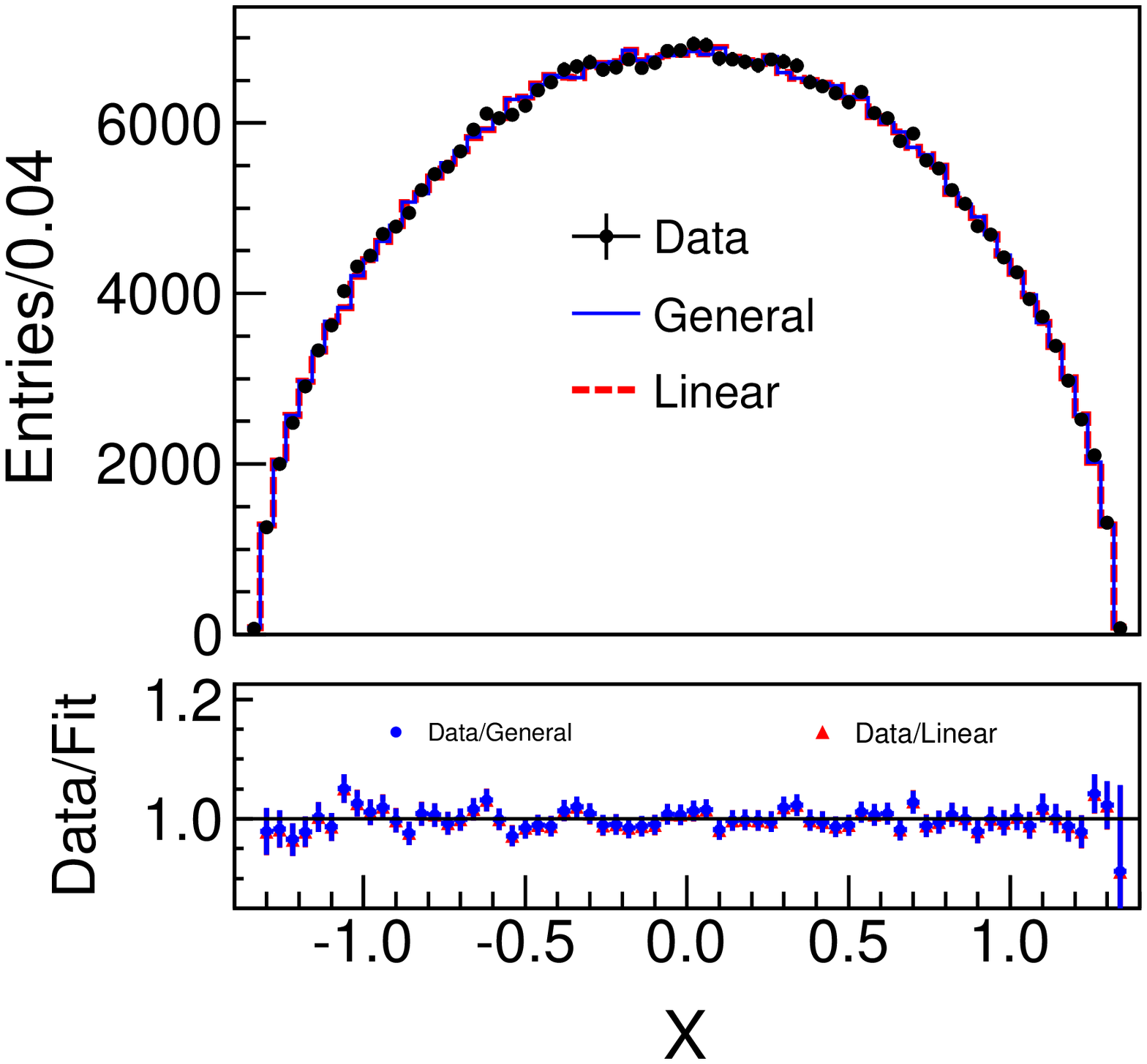}\put(-40,190){\bf (a)}
		\includegraphics[height=8cm]{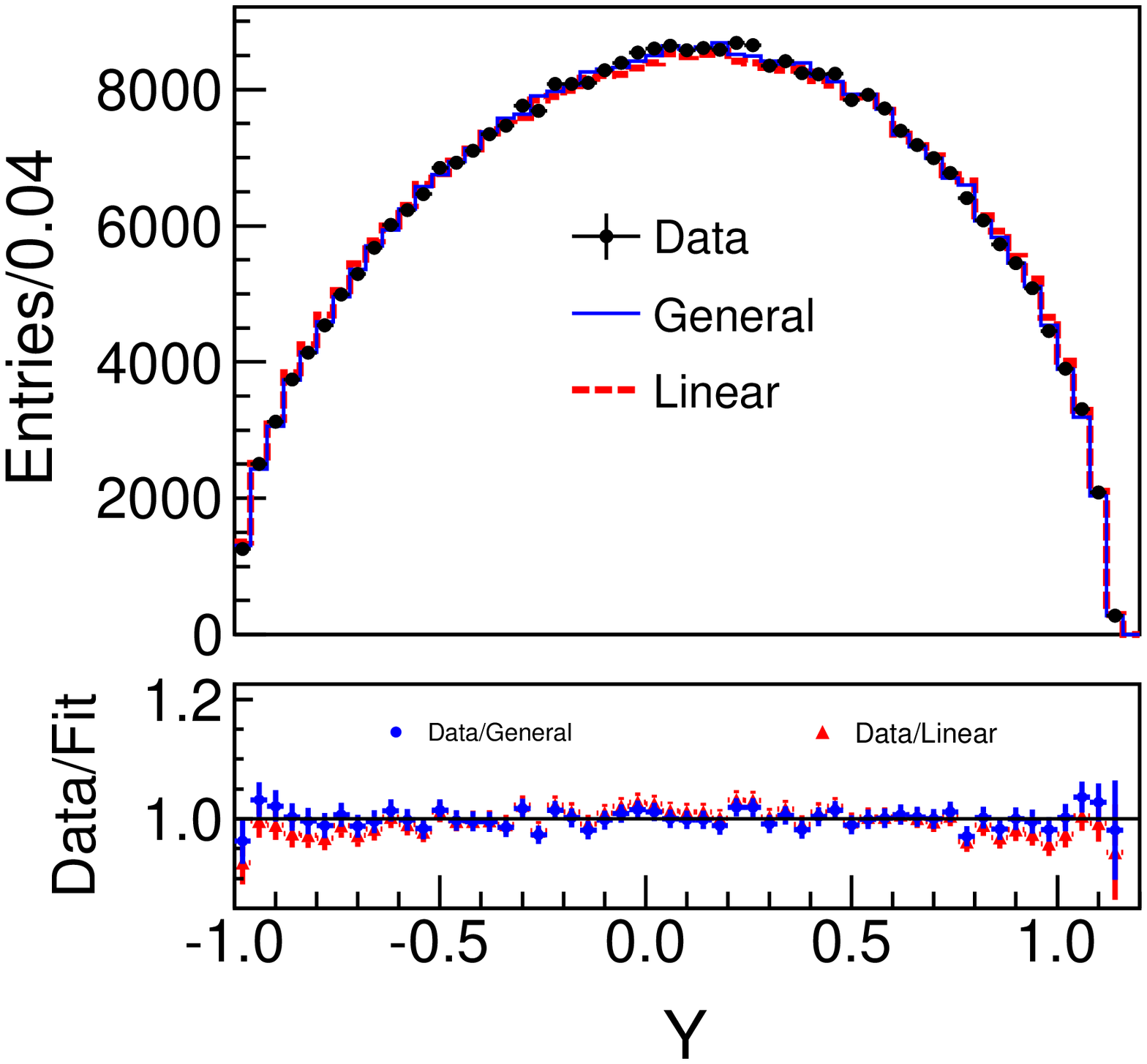}\put(-40,190){\bf (b)}
    \caption{\label{Fig:ChaFit} Projections of the fit results onto (a) $X$ and
		(b) $Y$ in the general
		(solid histograms) and linear (dashed histograms) representations for
		$\etap\ra\eta\pip\pim$, where the dots with error bars represent data.}
\end{figure*}

After the above requirements, 351,016 $\etap\ra\eta\pip\pim$ candidate events
are selected, with an averaged efficiency of 31.2\% and a background contribution of less than 0.3\%.
Figure~\ref{Fig:ChaDalitz} shows the Dalitz
plot in the variables $X$ and $Y$ for the selected events. The corresponding
projections on $X$ and $Y$ are shown as the dots with error bars in
Figs.~\ref{Fig:ChaFit} (a) and \ref{Fig:ChaFit}(b), respectively.
The resolution on the variables $X$ and $Y$ over
the entire kinematic region, determined from the MC simulation, are 0.03 and 0.02, respectively.

Unbinned maximum likelihood fits to the data are performed to determine
the free parameters in the decay amplitude squared [Eqs.~(\ref{eq:ampxy})
and (\ref{eq:ampxy_linear})].
To account for the resolution and detection efficiency, the amplitude
squared is convoluted with a function $\sigma(X,Y)$ parametrizing the
resolution and multiplied by a function $\varepsilon(X,Y)$
parametrizing the detection efficiency. Both functions are derived
from MC simulations. Two double Gaussian functions are used for
$\sigma(X,Y)$, while $\varepsilon(X,Y)$ is estimated as the average
efficiencies of local bins. With the normalization, one derives the
probability density function ${\cal P}(X,Y)$, which is applied in
the fit,
\begin{equation}\label{eq:pdf}
\begin{gathered}
  {\cal {P}}(X,Y) = \\
  \frac{|M(X,Y)|^{2}\otimes\sigma(X,Y)\cdot \varepsilon(X,Y)}{\int_{DP}{(|M(X,Y)|^{2}\otimes\sigma(X,Y)\cdot \varepsilon(X,Y))dXdY}}.
\end{gathered}
\end{equation}
The integral over the full Dalitz plot range
(DP) gives the normalization factor in the denominator.
The fit is done by minimizing the negative log-likelihood value
\begin{equation}
  -\ln {\cal L} = -\sum_{i=1}^{N_\text{event}}\ln{}{{\cal{P}}(X_{i},Y_{i})},
\end{equation}
where ${\cal P}(X_{i},Y_{i})$ is evaluated for an event $i$,
and the sum runs over all accepted events.

Imposing charge conjugation invariance by setting the coefficient of  odd powers in $X$ ($c$) to zero
in the general representation, the fit yields
following parameters:
\begin{equation} \label{eq:fitrescha}
\begin{matrix}
a & = & -0.056\pm0.004, \\
b & = & -0.049\pm0.006, \\
d & = & -0.063\pm0.004. \\
\end{matrix}
\end{equation}
Here, the uncertainties are statistical only. The corresponding correlation matrix
of the fit parameters is
\begin{equation}
\begin{pmatrix}
   & \vline &     b     &    d     \\\hline
 a & \vline &  -0.417   &       -0.239 \\
 b & \vline &           & \phantom{-} 0.292 \\
\end{pmatrix}
.
\end{equation}
Projections of the fit result on $X$ and $Y$ are illustrated as
the solid histograms in Fig.~\ref{Fig:ChaFit}.

To check for the
existence of a charge conjugation violating
term,
an alternative fit with the parameter $c$ free is performed.
The resultant value, $c=(2.7\pm2.4)\times10^{-3}$,
is consistent with zero.
Compared with the nominal fit results, the parameters
$a$, $b$ and $d$ are almost unchanged, and the statistical significance for
a nonzero value of the parameter $c$ is determined to be $0.7\sigma$ only.

Alternative fits including the extra terms $fY^3+gX^2Y$ or
$eXY+hXY^2+lX^3$ in the general representation are also performed, resulting in
$f=-0.004\pm0.012$, $g=0.008\pm0.010$ or
$e=0.005\pm0.007$, $h=0.004\pm0.006$, and $l=0.007\pm0.013$, respectively,
while the other parameters are unchanged.

A fit based on the linear representation is also performed and yields
the following values:
\begin{equation} \label{eq:fitres_linear_cha}
\begin{matrix}
\Re(\alpha) & = & -0.034\pm0.002, \\
\Im(\alpha) & = & \phantom{-}0.000\pm0.019, \\
d          & = & -0.053\pm0.004. \\
\end{matrix}
\end{equation}
The imaginary part of $\alpha$ is consistent with zero.  This can be
understood by the observation that the coefficient $b$ in the general
representation is negative.

Subsequently we will consider the
fit result with $\Im(\alpha)$ fixed at zero.
The parameters $\Re(\alpha)$ and $d$ and their uncertainties
remain the same as in Eq.~(\ref{eq:fitres_linear_cha}),
and the
correlation coefficient between $\Re(\alpha)$ and $d$ is $-0.137$.
The log-likelihood value is lower by 33.9 compared with the fit using the
general representation, which indicates
that the linear representation is less compatible
with the data.
Projections on $X$ and $Y$ based on this result are illustrated as the dashed histograms in
Fig.~\ref{Fig:ChaFit}. The presented residuals show that the fit is slightly worse
to describe the data in $Y$ projection comparing to the general one.

The potential charge conjugation violating is also checked in the linear representation
by performing an alternative fit with a free parameter $c$.
The resultant value, $c=(2.7\pm2.4)\times10^{-3}$,
is also consistent with zero, while the parameters
$\Re(\alpha)$ and $d$ are almost unchanged
compared with the nominal fit results.

\section{Measurement of the matrix element for the decay $\etap\ra\eta\pio\pio$}\label{sec:neutral}

In the reconstruction of $\jpsi\ra\gamma\etap$ with $\etap\ra\eta\pio\pio$
and $\eta/\pio\ra\gamma\gamma$, candidate events must have at least seven
photons and no charged track. The selection criteria for photon candidates
are the same as those for $\etap\ra\eta\pip\pim$, except that the requirement
on the angle between photon candidates and any charged track is not used. A requirement
of an EMC cluster timing with respect to the most energetic photon ($-500 \le T \le 500$ ns) is also
used. The photon with the largest energy in the event is assumed
to be the radiative photon originating from the $\jpsi$ decay. For the remaining
clusters, pairs of photons are combined into $\pio/\eta\ra\gamma\gamma$
candidates, which are subjected to a one-constraint (1C) kinematic fit by constraining the invariant mass
of the photon pair to be the nominal $\pio$ or $\eta$ mass.
The $\chisq$ for this 1C kinematic fit is required to be less than 25.
To suppress $\pio$ miscombinations, the $\pio$ decay angle $\theta_{\rm decay}$,
defined as the polar angle of one of the decay photons in the
$\gamma\gamma$ rest frame with respect to the $\pio$ flight direction,
is required to satisfy $|\cos\theta_{\rm decay}|<0.95$. Then an eight-constraint
(8C) kinematic fit is performed for the $\gamma\eta\pio\pio$ combination enforcing
energy-momentum conservation and constraining the invariant masses of the three photon pairs
and the $\eta\pio\pio$ combination to the nominal $\pio/\eta$ and
$\etap$ masses.
If more than
one combination is found in an event, only the one with the smallest
$\chi^{2}_{8C}$ is retained.  Events with $\chisq_{8C}<100$ are accepted
for further analysis.

\begin{figure}
    \centering
    \includegraphics[height=6.5cm]{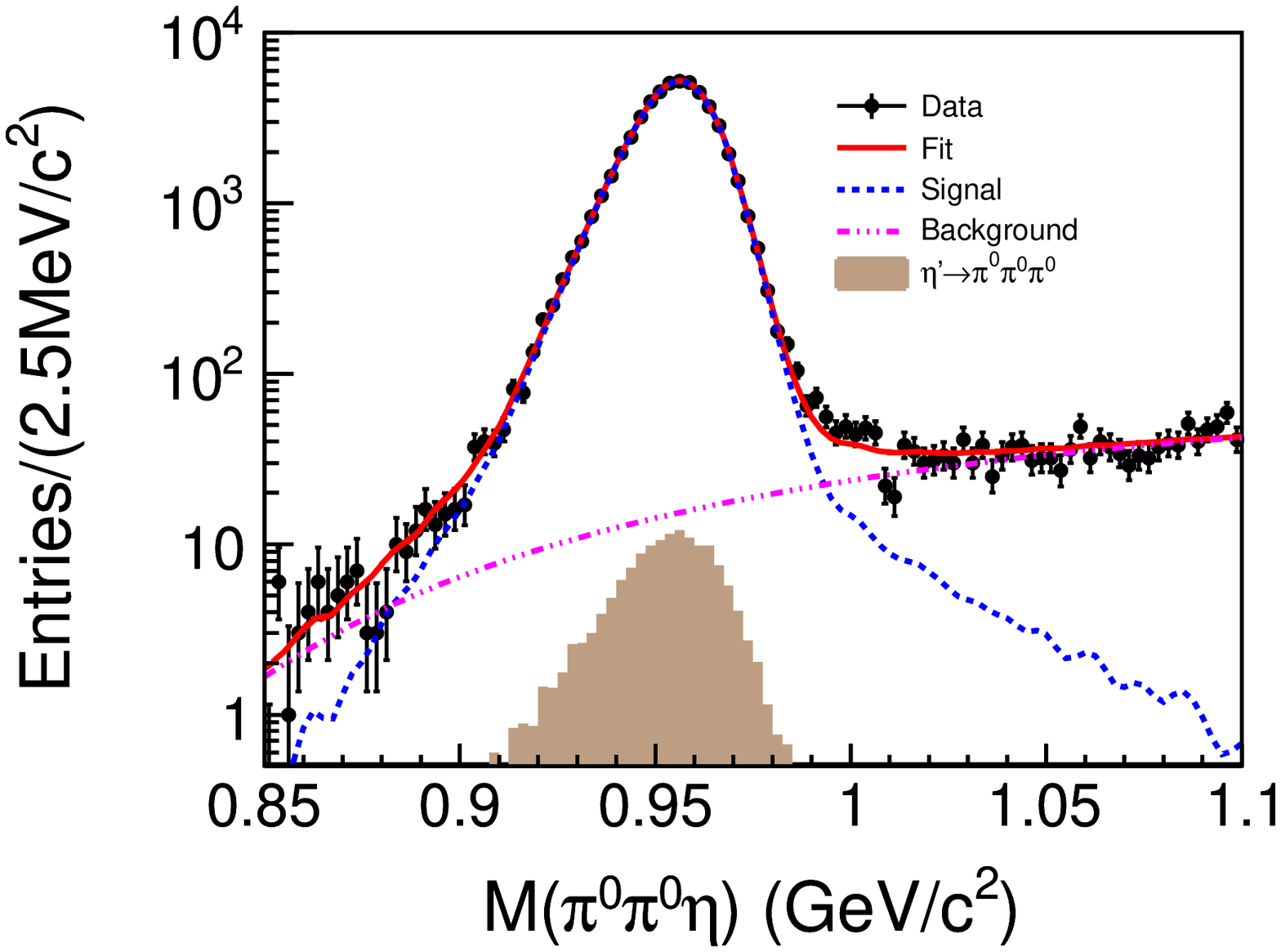}
    \caption{\label{fig:evtneu} Invariant mass spectrum of $\pio\pio\eta$ candidates
		 without $\eta$ and $\etap$ mass constraints applied in the kinematic fit,
		 and requiring the $\gamma\gamma$ invariant mass within the $\eta$ signal region.
		}
\end{figure}

To estimate the backgrounds, an alternative selection is performed where
$\eta$ and $\etap$ mass constraints in the kinematic fit are removed.
The resulting $\pio\pio\eta$ invariant mass spectrum is shown
in Fig.~\ref{fig:evtneu}, after requiring the $\gamma\gamma$ invariant
mass within the $\eta$ signal region, (0.518, 0.578)~GeV/$c^2$.
The inclusive MC study shows that
the surviving backgrounds mainly consist of the peaking background
$\etap\ra\pio\pio\pio$ and a flat contribution from $\jpsi\ra\omega\eta$
with $\omega\ra\gamma\pio$ and $\eta\ra\pio\pio\pio$.
From this MC sample, the background contamination is estimated to be about 0.9\%,
which is consistent with the estimation obtained from a 
fit to $M(\eta\pio\pio)$ and therefore neglected in the determination
of the Dalitz plot parameters.
In the fit, the signal is described by the MC simulated
shape convoluted with a Gaussian function representing the difference of
the mass resolution between the data and MC simulation.
The shape and the yield of the peaking
background $\etap\ra\pio\pio\pio$ are fixed according to the dedicated
MC simulation~\cite{PhysRevD.92.012014}. A third-order polynomial function
is used to represent the smooth background contribution.

\begin{figure}
	\includegraphics[height=6.5cm]{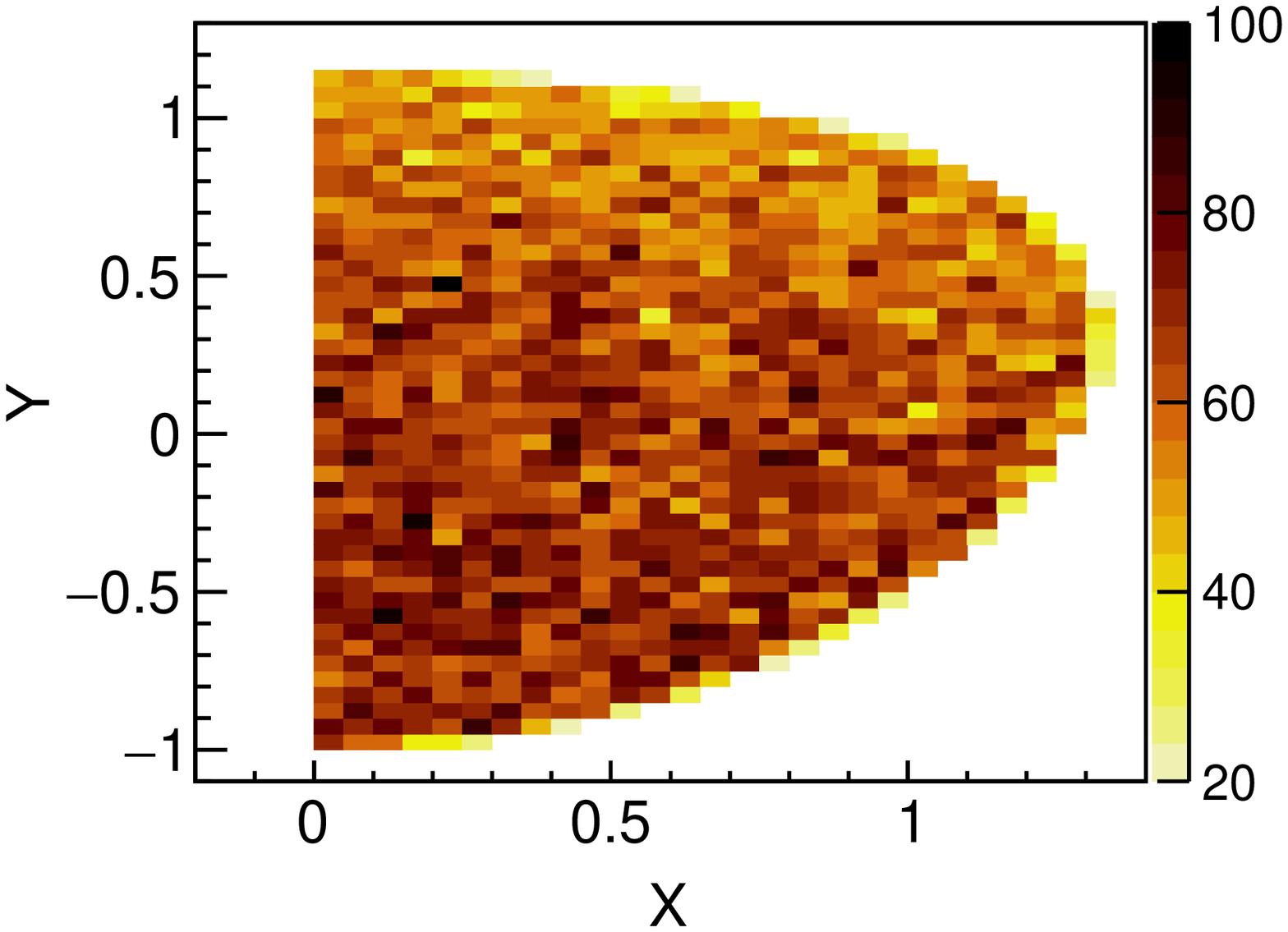}
  \caption{\label{fig:NeuDalitz} Dalitz plot for
	$\etap\ra\eta\pio\pio$ from data.}
\end{figure}

\begin{figure*}
	\includegraphics[height=8cm]{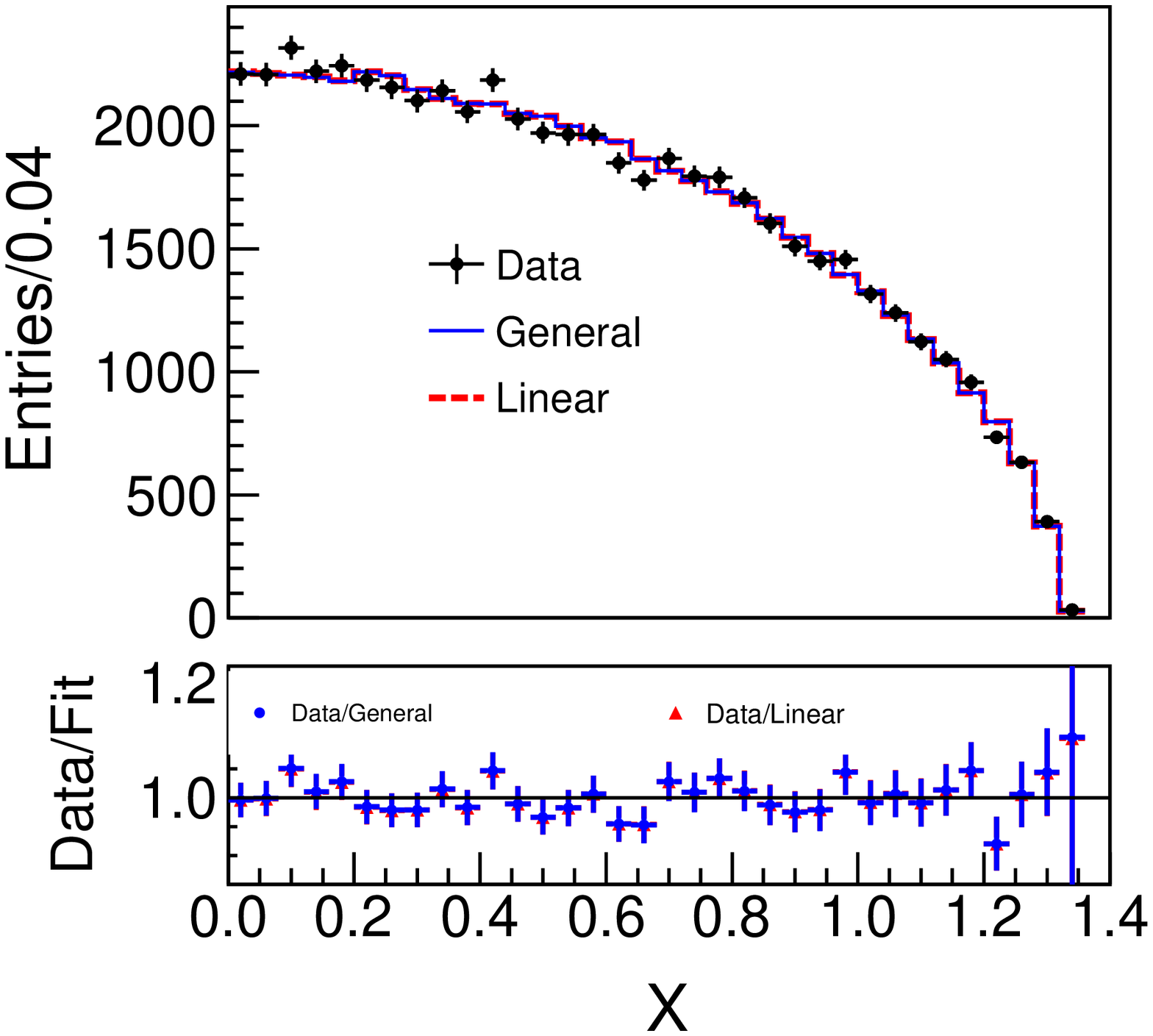}\put(-40,190){\bf (a)}
	\includegraphics[height=8cm]{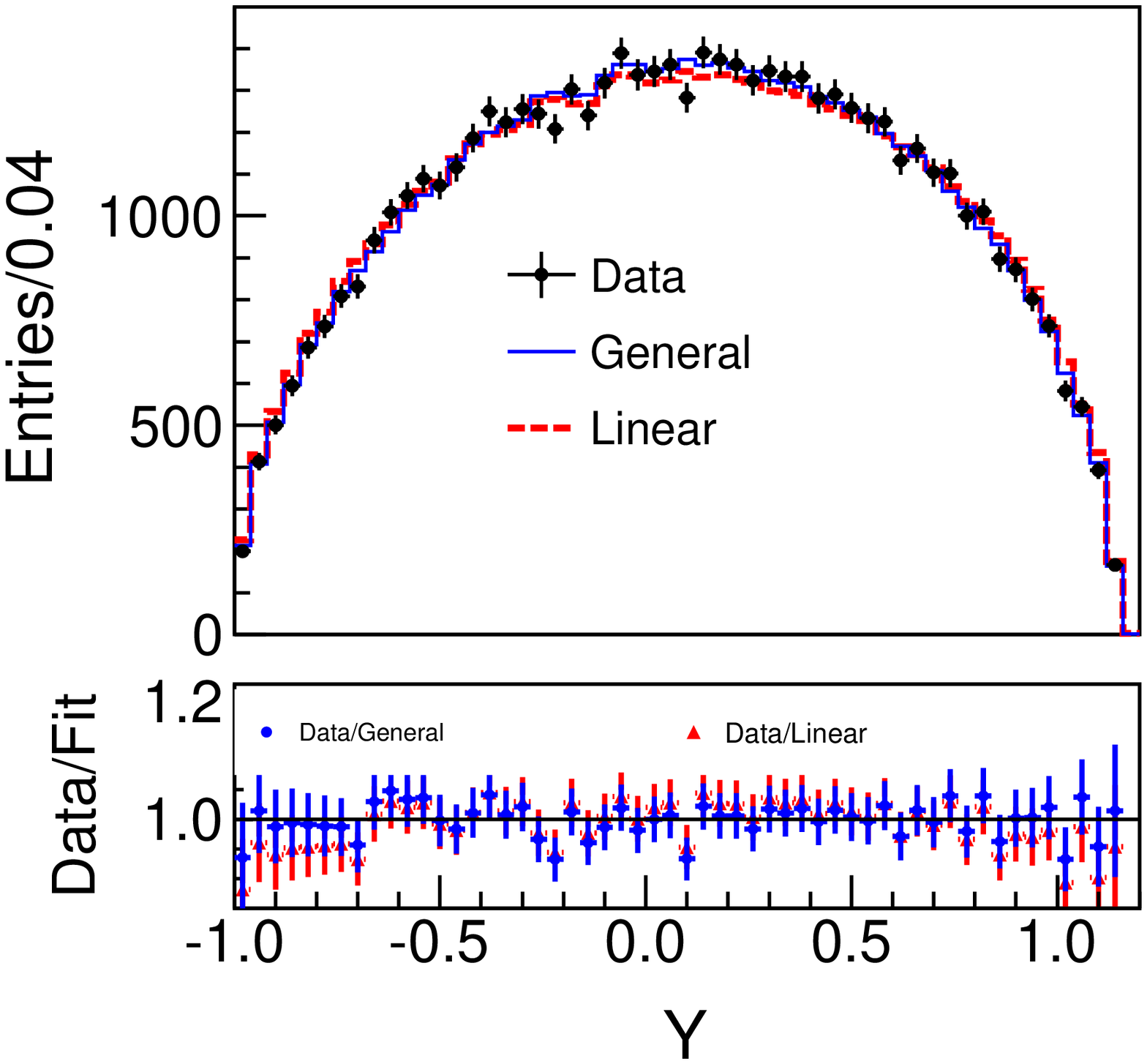}\put(-40,190){\bf (b)}
  \caption{\label{fig:NeuFit}	Projections of the fit results on (a) $X$ and
	(b) $Y$ in the general (solid histograms)
	 and linear (dashed histograms) representations for $\etap\ra\eta\pio\pio$,
	 where the dots with error bars represent data.}
\end{figure*}

After the above requirements, 56,249 $\etap\ra\eta\pio\pio$ candidate
events are selected, with an averaged efficiency of 9.6\% and a 0.9\% background level.
The Dalitz plot of selected events is displayed in Fig.~\ref{fig:NeuDalitz}.
The corresponding projections on $X$ and $Y$ are shown
as the dots with error bars in Fig.~\ref{fig:NeuFit}.
The resolution on $X$ and $Y$ over
the entire kinematic region, determined from the MC simulation, are 0.05 and 0.04, respectively.

As in the analysis of the $\etap\ra\eta\pip\pim$, an unbinned
maximum likelihood fit method is used to determine the Dalitz plot parameters.
The resolution is described with two double Gaussian functions, and
the detection efficiencies in different $X$ and $Y$ bins are obtained from the 
MC simulation.
From a dedicated study with the control sample
of $\jpsi\ra\pip\pim\pio$, we find that the reconstruction efficiency for
the $\pio$ candidate differs significantly between data and the MC simulation
at low $\pio$ momenta.
Thus, to describe the detection efficiency more accurately,
an efficiency correction depending on the $\pio$ momentum is carried out,
and the error of this correction will be considered in
systematic uncertainty.

Considering the strict constraint from the symmetry of the wave function, only
the fits without odd powers of $X$ are performed.
The fit based on the general representation yields the coefficient
(with statistical uncertainties only) and the corresponding correlation matrix,
\begin{equation}
	\begin{matrix}
		a  =  -0.087 \pm 0.009,  \\
		b  =  -0.073 \pm 0.014,  \\
		d  =  -0.074 \pm 0.009,
	\end{matrix}
\end{equation}
\begin{equation}
\begin{pmatrix}
   & \vline &      b   &    d   \\\hline
 a & \vline &   -0.495 & -0.273 \\
 b & \vline &   & \phantom{-} 0.273
\end{pmatrix}
.
\end{equation}
Similarly to the case $\etap\ra\eta\pip\pim$, the fit gives a negative value of the coefficient $b$.
The projections of the fit results on
$X$ and $Y$ are shown as the solid
histograms in Fig.~\ref{fig:NeuFit}.

Extra terms $fY^3$ and $gX^2Y$ in the general representation are also
added in an alternative fit, resulting in
$f=-0.023\pm0.028$ and $g=0.024\pm0.025$.
The significance for nonzero values of $f$ and $g$ is 0.6$\sigma$.

In the fit based on the linear representation,
the imaginary part of $\alpha$ (fitted to be $0.000 \pm 0.038$)
does not contribute to the fit quality,
as in the $\etap\ra\eta\pip\pim$ case.
Thus, the nominal fit omitting $\Im(\alpha)$ gives the results
(statistical uncertainties only),
\begin{equation}
	\begin{matrix}
		 \Re(\alpha) & = & -0.054 \pm 0.004,  \\
		     d      & = & -0.061 \pm 0.009.
	\end{matrix}
\end{equation}
The
correlation coefficient between the two parameters is $-0.170$.
Compared to the fit based on the general representation, the log-likelihood value
is reduced by 13.7.
Projections on $X$ and $Y$ are illustrated as
the dashed histograms in Fig.~\ref{fig:NeuFit}.
Again, the fits based on the two different representations give similar results for the
$X$ projections,
but slightly worse for $Y$ in the linear case.

\section{Systematic Uncertainties}

Various sources of systematic uncertainties on the measured Dalitz plot
parameters have been investigated, including tracking efficiency,
kinematic fit, efficiency correction,
and resolution. For the decay $\etap\ra\eta\pio\pio$, additional uncertainties
associated with photon miscombination, $\pio$ and $\eta$ reconstruction are
also considered.

Differences between the data and MC simulation for the tracking efficiency of
charged pions are investigated using the control sample $\jpsi\ra p {\bar p}\pip\pim$.
A momentum dependent correction on the detection efficiency is obtained by comparing
the efficiency between the data and MC simulation.
Similarly, a momentum dependent correction for the $\eta$
reconstruction efficiency is obtained with the control sample of $\jpsi\ra\gamma\eta\pip\pim$.
Then alternative fits are performed by incorporating the efficiency
corrections for charged pions or $\eta$. Changes of the Dalitz plot parameters
with respect to the nominal results are assigned as the systematic uncertainties.
A momentum-dependent $\pio$ reconstruction efficiency correction 
has been applied in the nominal fit; the associated systematic uncertainties are estimated
by changing the correction factor by one of its standard deviation and repeating the fit.
In comparison with $M(\pi^0\pi^0)$ without the $\pi^0$ reconstruction efficiency correction,
it is found that this correction has little impact on the cusp region.

The possible miscombination of photons in signal MC samples has been studied by
matching the generated photon pairs to the selected $\pio$ or $\eta$ candidates.
The fraction of events with wrong combinations is determined to be $2.7\%$
for $\etap\ra\eta\pio\pio$.
Alternative fits are performed to the MC simulated sample with only truth-tagged events and the
ones including miscombinations, individually.
The difference between those two results are taken as the systematic uncertainties.

To estimate the uncertainties associated with the kinematic fitting procedure,
the fit results are compared using a 4C (6C) instead of a 6C (8C) kinematic
fit for $\etap\ra\eta\pip\pim$ ($\etap\ra\eta\pio\pio$) and the corresponding
changes in the fit parameters are taken as systematic uncertainties.

To estimate the uncertainties associated with the efficiency correction
in Eq.~\ref{eq:pdf},
we change the Dalitz plot variables $X$ and $Y$ to the so-called square Dalitz plot
variables $M(\eta\pi)^{2}$ and $\cos\theta$, where $\theta$ is the angle
between the two pions in the rest frame of $\eta\pi$. Alternative fits
are performed with the efficiency correction based on the newly defined Dalitz plot
variable and the resultant changes of
the Dalitz plot parameters with respect to the nominal results are assigned
as systematic uncertainties.

To estimate the impact from the nonflat resolution in the $X-Y$ plane,
the biases from input/output checks are taken as the systematic
uncertainties. The impact from different resolutions of the Dalitz
plot variables between data and the MC simulation is estimated by alternative fits
varying the resolutions by $\pm10\%$. It is found that the
change of the results is negligible.
The effect of neglecting the residual background is checked by alternative fits
including MC simulated backgrounds and
found to be insignificant.

All of the above uncertainties are summarized in Table~\ref{syserr}.
Assuming all the sources of systematic uncertainty are independent, the total systematic uncertainties
for the Dalitz plot parameters are obtained by adding the individual values in quadrature,
shown in the last row of Table~\ref{syserr}.

\begin{table*}
\centering
 \caption{\label{syserr} Systematic uncertainties of the Dalitz plot parameters
 in the generalized and linear representations.
 }
 \begin{tabular}{c|ccc|cc|ccc|cc}\hline\hline
 \multirow{3}{*}{Source} & \multicolumn{5}{|c|}{$\etap\ra\eta\pip\pim$} & \multicolumn{5}{c}{$\etap\ra\eta\pio\pio$}\\\cline{2-11}
& \multicolumn{3}{|c|}{General representation} & \multicolumn{2}{c|}{Linear representation} & \multicolumn{3}{|c|}{General representation} & \multicolumn{2}{c}{Linear representation} \\
       								 & $a$ & $b$ & $d$ & $\Re(\alpha)$ & $d$  &	 $a$ & $b$ & $d$ & $\Re(\alpha)$ & $d$ \\\hline

Tracking efficiency    & 0.0018 & 0.0044 & 0.0021 & 0.0015 & 0.0013 & $...$  & $...$  & $...$  & $...$  &  $...$ 	\\
$\pio$ efficiency      & $...$  & $...$  & $...$  & $...$  & $...$ 	& 0.0006 & 0.0007 & 0.0001 & 0.0004 &  0.0003 \\
$\eta$ efficiency 		 & $...$  & $...$  & $...$  & $...$  & $...$ 	& 0.0012 & 0.0014 & 0.0001 & 0.0005 &  0.0003 \\
Photon miscombination  & $...$  & $...$  & $...$  & $...$  & $...$ 	& 0.0002 & 0.0024 & 0.0013 & 0.0004 &  0.0009 \\
Kinematic fit          & 0.0009 & 0.0035 & 0.0024 & 0.0007 & 0.0031 & 0.0041 & 0.0031 & 0.0019 & 0.0005 &  0.0016	\\
Efficiency presentation& 0.0009 & 0.0002 & 0.0007 & 0.0005 & 0.0007 & 0.0002 & 0.0005 & 0.0004 & 0.0002 &  0.0005	\\
Resolution  			     & 0.0006 & 0.0009 & 0.0004 & 0.0005 & 0.0015 & 0.0044 & 0.0021 & 0.0030 & 0.0004 &  0.0048	\\\hline
Total                  & 0.0023 & 0.0057 & 0.0033 & 0.0018 & 0.0038 & 0.0062 & 0.0047 & 0.0038 & 0.0010 &  0.0052	\\\hline\hline

 \end{tabular}
\end{table*}

\section{Comparison between $\etap\ra\eta\pip\pim$ and $\etap\ra\eta\pio\pio$ and
Search for cusp effect in $\etap\ra\eta\pio\pio$}\label{sec:cusp}

\begin{figure*}
	\includegraphics[height=6.5cm]{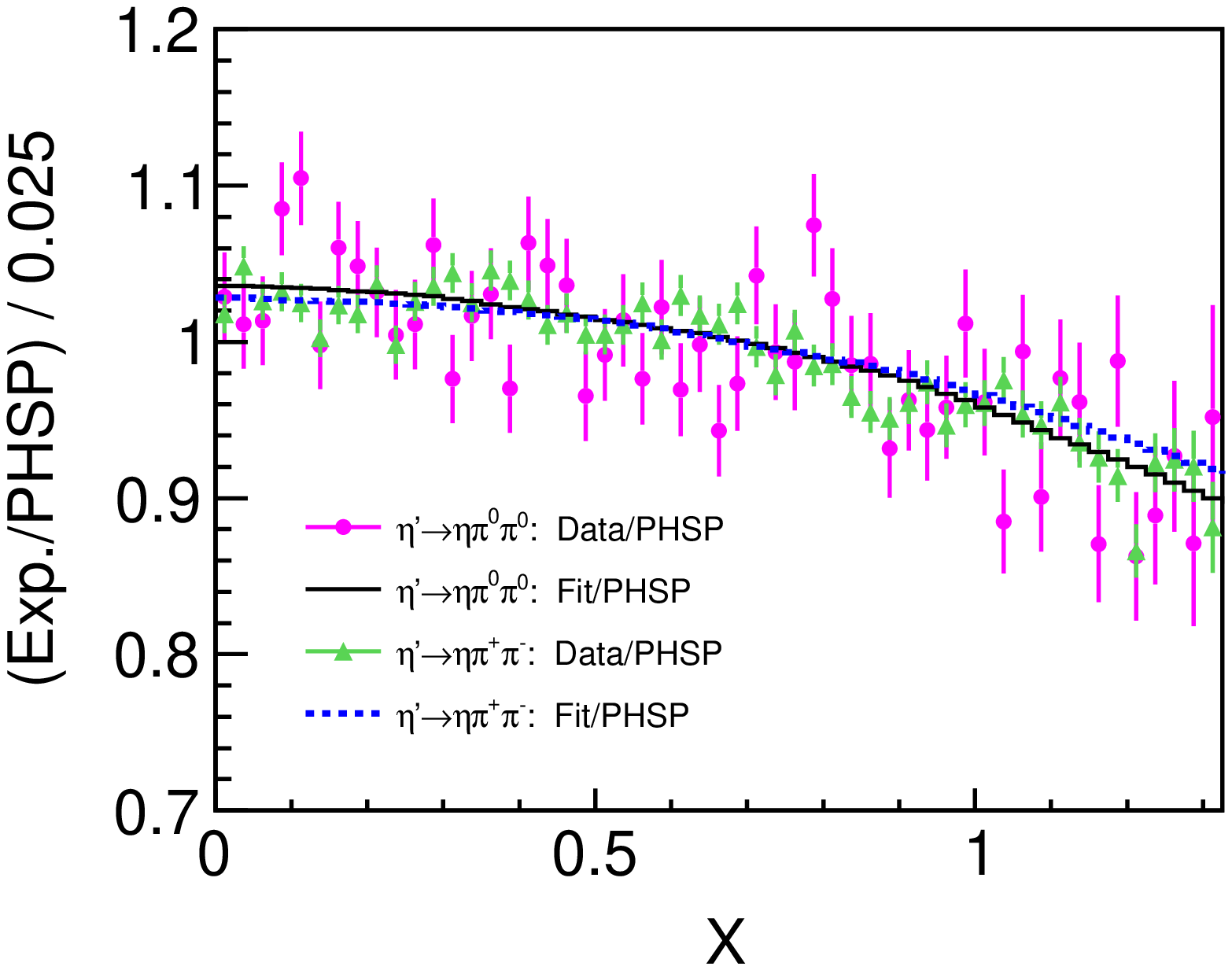}\put(-60,150){\bf (a)}
	\includegraphics[height=6.5cm]{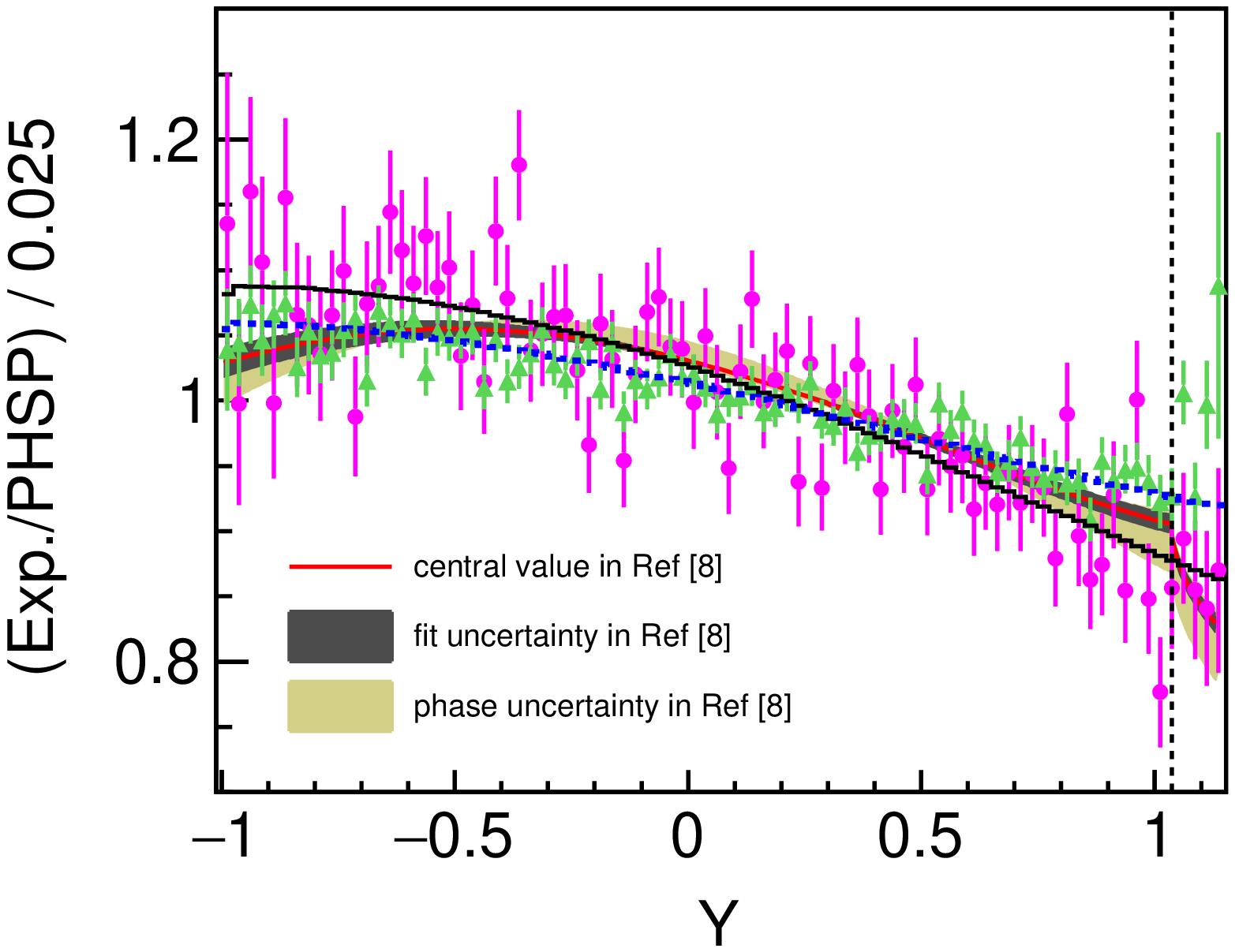}\put(-60,150){\bf (b)}

	\includegraphics[height=6.5cm]{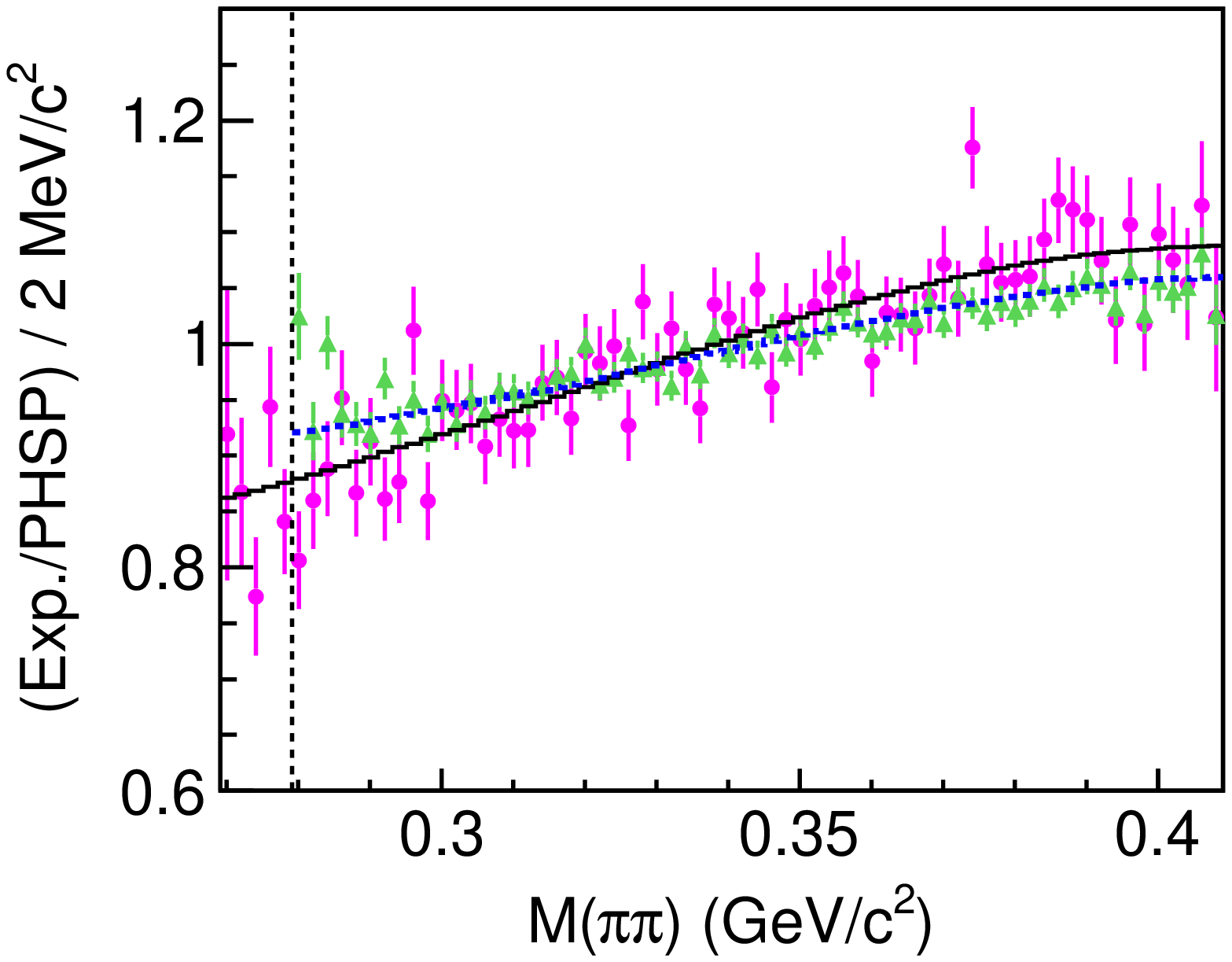}\put(-60,150){\bf (c)}
	\includegraphics[height=6.5cm]{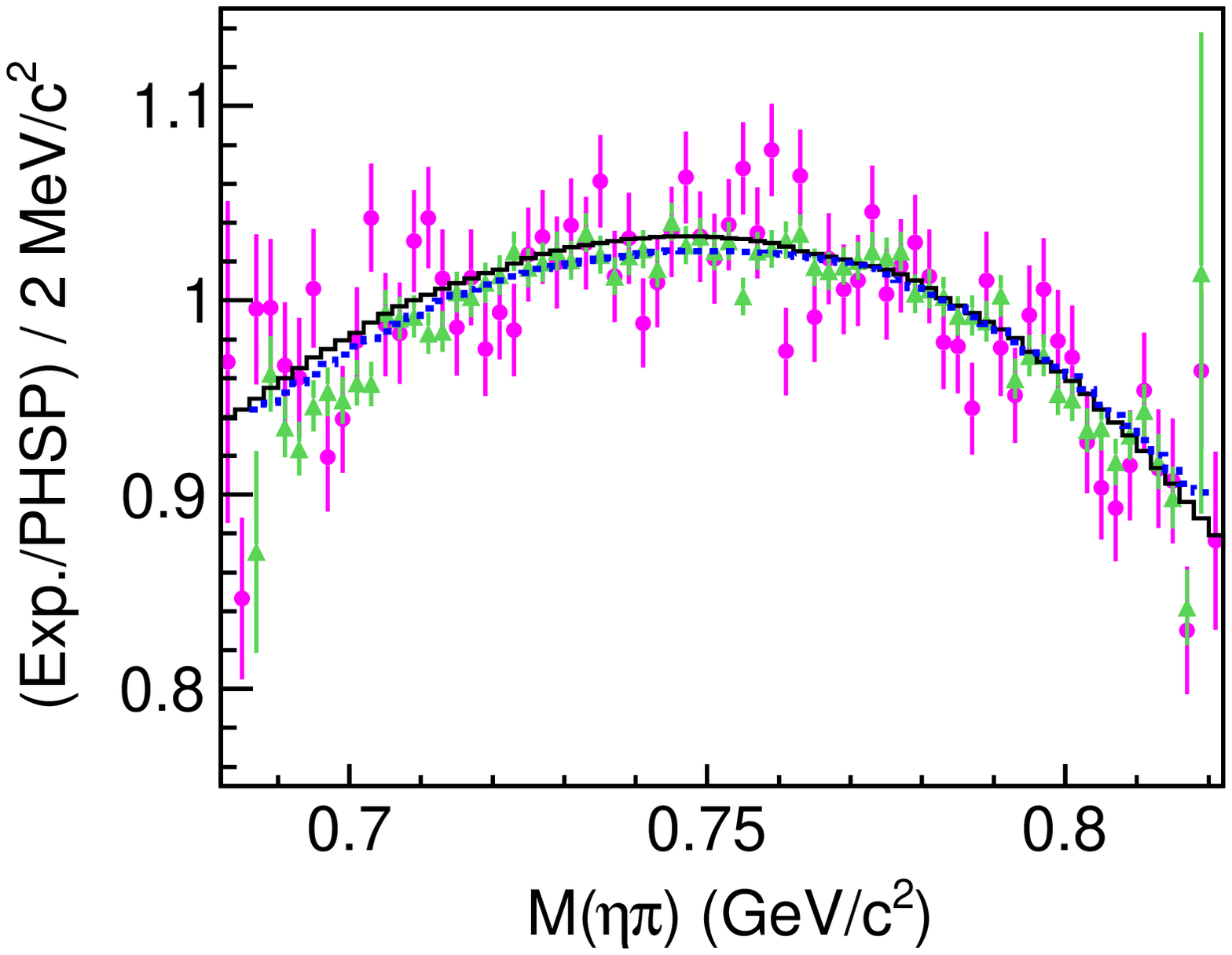}\put(-60,150){\bf (d)}
  \caption{\label{fig:twomode} The (a) $X$, (b) $Y$, (c) $M(\pi\pi)$, and
	(d) $M(\eta\pi)$ distributions of data (dots and triangles)
	and fit results in the general representation (histograms and dotted-lines) divided
	by the phase space distribution for $\etap\ra\eta\pip\pim$ and $\etap\ra\eta\pio\pio$.
	Overlaid on (b) is the prediction for $\etap\ra\eta\pio\pio$
	based on the previous BESIII fit result
	in Ref.~\cite{Isken2017} (smooth line),
	where the two error bands give the uncertainties
	resulting from the fit and originating
	from the variation of the phase input, respectively.
	The vertical lines in (b) and (c) correspond to the $\pip\pim$ mass threshold.}
\end{figure*}

After the event selection criteria presented in Secs.~\ref{sec:charge}
and~\ref{sec:neutral}, clean $\etap\ra\eta\pip\pim$ and $\etap\ra\eta\pio\pio$ 
samples are selected.
A comparison between the charged and neutral decay modes could be performed
by dividing the acceptance corrected experimental distributions with the
corresponding phase space distributions
on variables $X$ ( absolute value for $\etap\ra\eta\pip\pim$), $Y$,
$M(\pi\pi)$, and $M(\eta\pi)$, 
which are shown in Fig.~\ref{fig:twomode},
together with the Dalitz plot fit results based on the general representation.
Although the statistical errors are large, the trends of the experimental distributions
on $Y$ and $M(\pi\pi)$
between the charged and neutral mode are obviously different,as the high statistical simulation based
on the fit results on the general representation shows.
At the same time, the difference on $X$ and $M(\eta\pi)$ are smaller.
The observed differences are likely to be related to the $\pi\pi$ and $\eta\pi$ final interaction.

The ratio between experimental and phase space distributions on
$Y$ and $M(\pi\pi)$, Figs.~\ref{fig:twomode} (b) and \ref{fig:twomode}(c),
also provide the possibility to check the cusp effect.
Overlaid on Fig.~\ref{fig:twomode} (b)
is the prediction for $\etap\ra\eta\pio\pio$ in Ref.~\cite{Isken2017} based on 
the previous BESIII fit result for
$\etap\ra\eta\pip\pim$~\cite{PRD83012003},
which are consistent with the experimental distribution within
statistical errors.
However, with current statistics, it is difficult to establish
the structure (cusp effect) near the $\pip\pim$ mass threshold.

\section{Summary}

With a sample of $1.31\times 10^9$ $J/\psi$ events collected with the BESIII detector,
clean samples of 351,016 $\eta^\prime\rightarrow\eta\pi^+\pi^-$ events
and 56,249 $\eta^\prime\rightarrow\eta\pi^0\pi^0$ events are selected
from $J/\psi$ radiative decays. Then the most precise measurements
of the matrix element for the $\eta^\prime\rightarrow\eta\pi^+\pi^-$ and
$\eta^\prime\rightarrow\eta\pi^0\pi^0$ decays as well as a search for
the cusp effect in $\eta^\prime\rightarrow\eta\pi^0\pi^0$ are performed.

Both the general and the linear representations
are used to determine the Dalitz plot parameters and
the corresponding results are summarized in Table~\ref{fitres}
including the systematic uncertainties.
The Dalitz plot parameters for both decays are in
reasonable agreement and more precise than the previous
measurements~\cite{PRL8426,Dorofeev200722, Blik2009231, PRD83012003}.
The results for $\eta^\prime\rightarrow\eta\pi^+\pi^-$ supersede
the previous BESIII measurement~\cite{PRD83012003}, which used a subsample
of the present data. As reported in Ref.~\cite{PRD83012003}, the discrepancy of
the parameter $a$ for $\eta^\prime\rightarrow\eta\pi^+\pi^-$ with respect
to the VES value~\cite{Dorofeev200722} is evident, which, at present,
stands at about 3.8 standard deviations. The values of the parameter $c$ in
$\eta^\prime\rightarrow\eta\pi^+\pi^-$ are all consistent with zero
within one standard deviation in both representations,
in agreement with the charge conjugation conservation in the strong interaction.
In addition, a discrepancy of 2.6 standard deviations for the parameter $a$ 
is observed between $\etap\ra\eta\pip\pim$ and $\etap\ra\eta\pio\pio$ processes,
indicating as isospin violation.
However, the result is not statistically significant enough to firmly establish
such a violation, and additional effects,
$e.g.$, radiative corrections~\cite{Kubis:2009sb},
should be considered in
future experimental and theoretical studies.

A comparison between the results obtained from the general representation and 
the theoretical predictions within the framework of U(3) chiral effective
field theory (EFT) incorporating with a relativistic coupled-channels approach~\cite{Borasoy2005383}
is given in Table~\ref{fitres}. 
In general,
our results are compatible with the theoretical expectations.
However, the theoretical prediction for the parameter $a$ from
$\eta^\prime\rightarrow\eta\pi^+\pi^-$ is about 2 times larger
than our result, and
the discrepancies on the parameter $d$ for both $\eta^\prime\rightarrow\eta\pi^+\pi^-$
and $\eta^\prime\rightarrow\eta\pi^0\pi^0$ are about
four standard deviations. 
Table~\ref{fitres} also provides the predictions obtained in the frameworks of large-N$_C$ ChPT and
resonance chiral theory (RChT) with the parameters $a$ fixed according to the boundaries measured in
Refs.~\cite{Dorofeev200722, Blik2009231}.
The expected values are consistent with our results within two standard deviations in both decay modes,
except that the parameter $d$ in $\etap\ra\eta\pip\pim$ is 3.1 standard deviations from the large-N$_C$ ChPT,
and the parameter $b$ in $\etap\ra\eta\pio\pio$ is 2.7 standard deviations from the RChT.

As previously mentioned, the linear and general representations are equivalent
for the case of $b>a^2/4$. However, the coefficients $b$ for the $Y^2$ term are
negative with 5.8 and 4.9 standard deviations to zero for $\etap\ra\eta\pip\pim$ and
$\etap\ra\eta\pio\pio$, respectively, which implies that these two
representations can not provide an identical description of data.
In case of the linear representation,
the results are in agreement with previous measurements and also provide
a reasonable description on the $X$ projection for both decay modes. However,
the goodness of fit on the $Y$ projections
are worse than
the general one. This is consistent with the conclusion reported
by the VES Collaboration~\cite{Dorofeev200722} that the linear
representation can not describe the data well.

We also attempt to search for the cusp effect in the decay
$\eta^\prime\ra\eta\pi^0\pi^0$. Inspection of the $\pi^0\pi^0$ mass spectrum
around the $\pip\pim$ mass threshold does not show evidence of a cusp
with current statistics.

\begin{table*}
\centering
 \caption{\label{fitres} Experimental and theoretical values of the Dalitz plot
 parameters for $\etap\ra\eta\pip\pim$ and $\etap\ra\eta\pio\pio$.
 The values for parameter $c$ and $\Im(\alpha)$ are given for comparison with previous experiments.}
% \resizebox{\textwidth}{!}{%
 \begin{tabular}{c|c|c|c|c|c|c|c|c}\hline\hline
  Para- & \multicolumn{5}{|c|}{$\etap\ra\eta\pip\pim$} & \multicolumn{3}{|c}{$\etap\ra\eta\pio\pio$}  \bigstrut\\\cline{2-9}
   meter & EFT~\cite{Borasoy2005383} & Large N$_C$~\cite{Escribano2011094} & RChT~\cite{Escribano2011094} & VES~\cite{Dorofeev200722} & This work &
	 EFT~\cite{Borasoy2005383} & GAMS-4$\pi$~\cite{Blik2009231} & This work \\\hline\hline

  $a$ & $-0.116(11)$ & \multicolumn{2}{|c|}{$-0.098(48)$ (fixed)} & $-0.127(18)$ & $-0.056(4)(2)$ & $-0.127(9)$ & $-0.067(16)$ & $-0.087(9)(6)$ \\\hline
  $b$ & $-0.042(34)$ & $-0.050(1)$ & $-0.033(1)$ & $-0.106(32)$ & $-0.049(6)(6)$ & $-0.049(36)$ & $-0.064(29)$ & $-0.073(14)(5)$ \\\hline
  $c$ & $...$ 		   & $...$ 	      & $...$    		 & $+0.015(18)$ & $0.0027(24)(18)$ & $...$ & $...$ & $...$ \\\hline
  $d$ & $+0.010(19)$ & $-0.092(8)$ & $-0.072(1)$ & $-0.082(19)$ & $-0.063(4)(3)$ & $+0.011(21)$ & $-0.067(20)$ & $-0.074(9)(4)$ \\\hline\hline

  $\Re(\alpha)$ & $...$ & $...$ & $...$ & $-0.072(14)$ & $-0.034(2)(2)$ & $...$ & $-0.042(8)$ & $-0.054(4)(1)$ \\\hline
  $\Im(\alpha)$ & $...$ & $...$ & $...$ & $0.000(100)$ & $0.000(19)(1)$ & $...$ & $0.000(70)$ & $0.000(38)(2)$ \\\hline
  $c$           & $...$ & $...$ & $...$ & $+0.020(19)$ & $0.0027(24)(15)$ & $...$ &  $...$  &  $...$  \\\hline
  $d$           & $...$ & $...$ & $...$ & $-0.066(34)$ & $-0.053(4)(4)$ & $...$ & $-0.054(19)$ & $-0.061(9)(5)$ \\
  \hline\hline
	\end{tabular}
%	}%
\end{table*}